# Use Factorial Design To Improve Experimental Reproducibility

By Bert Gunter, Retired

## Abstract

Systematic differences in experimental materials, methods, measurements, and data handling between labs, over time, and among personnel can sabotage experimental reproducibility. Uncovering such differences can be difficult and time consuming. Unfortunately, it is made more so when scientists employ traditional experimental procedures to explore possible sources of systematic variability by sequentially changing them one at a time to determine the magnitude of their effects. We use two simple examples to show how and why well known methods of factorial experimentation in which multiple potential sources are simultaneously varied provide a better alternative, can be understood as a straightforward extension of standard practice, and could be embedded into the quality control procedures of routine experimental practice. Doing so, we argue, would help mitigate at least some of the problems fueling the current "reproducibility crisis" roiling some scientific disciplines.

# Introduction

Experimental reproducibility, the idea that an observable experimental result can in some sense be replicated by anyone, is a bedrock tenet of science. It expresses the view that science is supposed to characterize the underlying behavior of a universal "reality," unlike, say, sports team preferences or art, where "truths" are individual and personal. Beauty may be in the eye of the beholder, but everyone is supposed to be able to behold science.

Nevertheless, translating this vague concept into an actionable definition of when an experiment or study has or has not been reproduced can be difficult (Saey, 2015; Baker, 2016). This is especially the case in disciplines like the social and biological sciences, where results tend to be inherently "noisy."

As an example, in a fascinating and readable account, Lithgow, Driscoll, and Phillips (2017) describe their efforts to find why consistent results from experiments in one lab on the effect of a druglike molecule on the lifetime of *C.elegans* nematode worms could not consistently be reproduced in another lab. They found many seemingly small differences between the labs that affected worm life span: food and reagent batches, handling techniques, types of incubators, small temperature fluctuations within incubators and lab environments, exact definitions and protocols for measurements, and genetic variation both within and between the species used, among others. In their words, "we have found how crucial it is to understand sources of variability between labs and experiments." They report spending over a year to expose these sources of variability and standardize the protocols, thereby "almost completely eliminating systematic differences in worm survival across our labs."

As they noted, biological experiments like theirs tend to be "fragile" – results are sensitive to many possibly subtle experimental details that may be unknown or undocumented and therefore likely to vary in different labs, thereby contributing to different results. Perhaps counterintuitively, some of this fragility is due to efforts to standardize as much as possible the materials, methods, measurements, and data analytical procedures *within* labs. Such standardization is of course intended to reduce variation in results *within a lab* to allow the effects of interest to be more precisely measured. The problem is that this increased precision can come at the cost of decreased generalizability: the results may reflect to an unknown extent the particular choices of materials and protocols, so that other labs using different materials and protocols may be unable to reproduce them.

For this reason, some researchers have suggested that less standardization and even deliberate variation is desirable to produce more robust results that are consistent across labs and time (Richter et. al., 2010). While many might consider this iconoclastic, even Lithgow et. al. recommended that researchers experiment with many more worms of several different species because of the systematic differences they found among them.

What may be surprising is that deliberately changing experimental methods or materials to investigate their effects is not a new idea: over 45 years ago, Jack Youden, a statistician working at the then National Bureau of Standards (now the National Institute for Science and Technology, NIST) gave examples of systematic errors in what would seem to be one of the essential activities of science: the determination of fundamental physical constants (e.g. the speed of light, the astronomical unit, etc.). In Youden (1972), he specifically noted:

> *The experimental method under consideration rests on the stand that the disagreement between investigators is largely a consequence of inadequate knowledge of the systematic errors inherent in the apparatus or equipment. We may state this another way. If we have accurate knowledge of the magnitude of the systematic errors in the components of the equipment much of the discrepancy among results would obviously be accounted for.*

Perhaps even more prescient:

> *What seems most interesting to me is that when measuring a physical constant usually no changes are made in the variables, or at most in one or two ... Equally clear is the fact that when another investigator makes his effort **every** component in the measuring system is changed. Reliance is again placed upon calibration and various corrections. Nothing is done about the fact that the results obtained by different investigators disagree a great deal more than would be expected in view of the 'paper' estimates of systematic errors used by the investigators. It appears to be an all or none situation. Everything gets changed in another laboratory. Almost nothing gets changed within a laboratory. This makes it impossible to single out and measure the sources of systematic errors.*

Few would disagree with the desirability of sorting out possible sources of systematic error, but the problem is how to do it. The nematode worm story shows how time consuming and tedious it can be, so this would not seem to be a model that could be widely applied.

Fortunately, there is an alternative, which Youden described in his paper: the use of factorial designs to efficiently investigate the possible effects of multiple experimental factors. The roots of the ideas are much older, however: they go back to the pioneering work of British geneticist and statistician R.A. Fisher at the Rothamsted agricultural experiment station in the 1920's (Joan Fisher Box, 1980).

Factorial designs are a standard part of experimental design methodology. They have been widely, if sporadically, applied in industrial, animal, psychological, and agricultural field experimentation (Festing, 2014; ILAR Journal, 2014; Snedecor & Cochran, 1989; Winer et. al., 1991; Daniel, 1976). Nevertheless, many researchers still seem unaware of or uncomfortable with these methods. As an attempt to redress this situation, we offer here a brief, conceptual exposition and examples of how to use factorial design for improving experimental robustness and reproducibility. Our intent is to show that, far from being arcane, basic factorial design can be understood as a straightforward extension of the experimental principles and practices that scientists routinely employ. While more advanced techniques of course require a deeper treatment of the mathematical underpinnings, the motivation is still the same: obtain as much information as possible within the limitations of available experimental resources.

## Using factorial designs to improve robustness and reproducibility

A standard scientific dogma is that to investigate the possible effects of multiple variables on a phenomenon of interest, only one at a time should be changed while holding the others constant, which we shall refer to as *OFAT*, for **O**ne **F**actor **A**t a **T**ime experimentation. In this way the effect of the single changing factor can be isolated without risk of confusing it with that of others. Fisher proved that this mantra is decisively wrong: the best way to determine the effects of multiple factors on experimental outcomes is to *vary them all simultaneously* in factorial designs, not one at a time! Nearly a hundred years of subsequent experience have shown that that he was right not only in theory, but also in practice.

In the present context, the factors of interest are the experimental materials, methods, equipment, environments, and so forth – the systematic differences between labs of which Youden (and Lithgow et al.) wrote. Of course, the all important specification of these factors will vary from discipline to discipline and experiment to experiment, to be determined by the investigators involved. But the factorial design structure itself can usually be kept fairly simple and standard.

Before going further, it is worth explicitly stating that factorial design is more than just a bunch of methods. In many respects, it represents a different paradigm for scientific experimentation in which a whole carefully pre-specified set of conditions are planned and investigated *together* instead of examining individual experimental results sequentially to determine the next condition at which to experiment based on the previous ones. While a detailed exposition of the framework cannot be provided here, Box, Hunter, and Hunter (2005) and Box and Draper (2007) are standard resources that include especially nice conceptual discussions of how the techniques fit within a paradigm of science as an iterative exploratory/confirmatory learning process. Coleman and Gunter (2014) present a brief nontechnical exposition of a small but useful subset of these ideas that shows how they can be efficiently implemented without extensive statistical machinery.

To gain insight into how factorial design can be used to ferret out systematic sources of experimental variability, we present a very simple, artificial example in which only two possible sources of variability are of concern. In practice, it is straightforward to include up to ten or more. An extended example with five variables that follows shows how this can be done. The limiting consideration is the ability to manage the experimental complexity, not the number of experimental "runs" – different combinations of settings – required. This point is discussed further in the Summary.

## The basic example

Suppose researchers want to assess the possible neurological effects of low concentrations of organic molecules that are widely distributed in the environment, e.g. from the breakdown of consumer goods such as packaging or plastics. To do this, they have developed a neuronal cell culture system into which agents can be added and subsequent changes in concentrations of key neural molecules measured.

In the case of one particular organic molecule, M, a collaborator at lab B has been unable to reproduce results obtained at lab A for this molecule. Lab *A* decides to investigate two possible sources of this failure: incubation temperature ( Lab A uses a different model incubator than Lab B for which the same temperature settings might differ somewhat between the labs); and batch to batch variation of a critical reagent *Q*, a component of the cell culture medium (the labs get different batches from the supplier).

Initially, Lab *A* proposes to investigate the effects of these two factors using the following OFAT protocol:

1) Incubate two flasks of *M* at the nominal temperature; the culture media in each flask will be made with different batches of *Q*, batches 1 and 2.
2) Incubate a flask of M with batch 1 of Q, but at an incubation temperature 2 °C higher than nominal.
3) Incubate a flask of M with batch 1 of Q, but an an incubation temperature 2 °C lower than nominal.

Step 1 allows the effect of Q batch to be determined while holding incubation temperature fixed. Steps (2) and (3) allow for the examination of different incubation temperatures while holding the *Q* batch fixed.  As a control, all steps are repeated with flasks without *M*. As will be shown below, this permits the experimenters to simultaneously assess the effect of M.

This experimental protocol can be more simply shown in the following table, where −, 0, + for temperature means low, nominal, and high temperature; + and − for M means media made with and without M, respectively. The Run ID column is simply a convenient way to notate each experimental combination of settings – i.e. each experimental *run*.

## Table 1:  Lab *A*'s OFAT Design Settings

| Run ID | M | Q batch | Incubation temperature |
|--------|---|---------|------------------------|
| 1 | + | 1 | 0 |
| 2 | + | 2 | 0 |
| 3 | + | 1 | - |
| 4 | + | 1 | + |
| 5 | - | 1 | 0 |
| 6 | - | 2 | 0 |
| 7 | - | 1 | - |
| 8 | - | 1 | + |

This simple tabular representation of the design is an indication of the underlying abstractions of experimental design that allow it to be studied statistically and from which its utility derives. There is another cognate geometric representation that provides additional insight. To show how it works, coordinate systems must be introduced for the three experimental variables, batch, temperature, and organic. For temperature, one could use the actual temperatures in whatever scale they are measured, but it is convenient and completely equivalent to simply linearly transform them so that −, 0, and + become −1, 0, and +1. We do the same for organic, although now this is completely artificial, as presence/absence of M has no meaningful ordering. [1]

One can now think of the design geometrically as a pattern in 3-d space -- a cube in fact -- where the X,Y, and Z axes are batch, temperature, and presence/absence of M. This can be shown as a pseudo 3-d perspective plot, which can be "flattened" to two 2-d "cubes", i.e. squares, one for "without M" (M = −1) and one for "with M" (M= +1).

---

[1] Although, of course, one could use M concentration as a scale, especially if several different concentrations were being tested, in which case this might be the best way to study M.

## Figure 1: OFAT Design Geometry

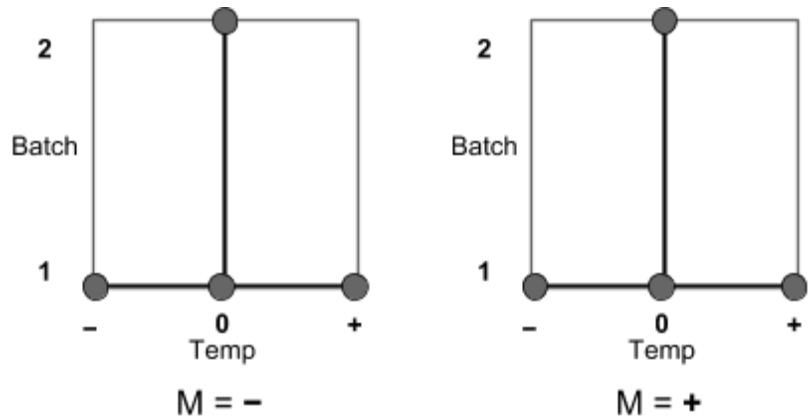

The 8 settings of the table are the points shown on the squares. For example, the lower right hand points in the squares corresponds to ID's 4 and 8 in the table.

Figure 1 illustrates the traditional OFAT (one-factor-at-a-time) approach. In contrast, Figure 2 shows what the factorial design approach looks like:

## Figure 2: Factorial Design Geometry

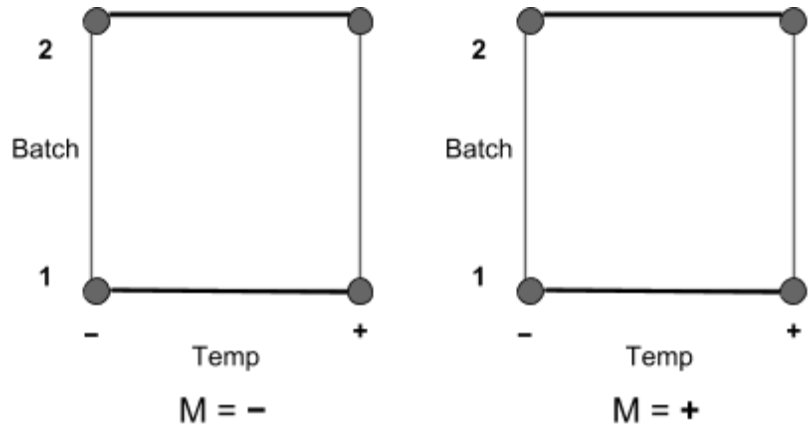

And here is the corresponding design table (in which Run ID's have been changed to avoid confusion with the previous table):

## Table 2: Factorial Design Settings

| Run ID | M | Q Batch | Incubation Temperature |
|--------|---|---------|------------------------|
| 9      | + | 1       | −                      |
| 10     | + | 2       | −                      |
| 11     | + | 1       | +                      |
| 12     | + | 2       | +                      |
| 13     | − | 1       | −                      |
| 14     | − | 2       | −                      |
| 15     | − | 1       | +                      |
| 16     | − | 2       | +                      |

The lower right hand corners now correspond to ID's 11 and 15.

Important note: the design is listed in this run order as an aid to understanding. In practice, it would be important to randomize the actual order in which the conditions were run as much as possible.

As both designs use the same number of runs, 8, what then is the advantage of the factorial design? To answer this, first, consider determining the effect of the different batches. For the OFAT design, the batch effect would be obtained as the differences in the neural molecule concentrations of the cultures between the two batches at the nominal temperature, both with and without M. These are the two top − bottom differences in concentrations on the vertical lines in the OFAT diagram. Their average, *(diff_with + diff_without)/2* is obviously an estimate of the average batch effect; but their difference, *(diff_with − diff_without)/2* is also of interest (the divisor of 2 comes from the underlying math, and can be thought of as providing a certain technical consistency). It estimates the *batch*M interactive* effect, which addresses the question: does the batch difference effect depend on whether M is or is not present? – or, equivalently, does the effect of M differ depending on which batch of Q is used (the details of this equivalence are omitted, but they are just simple arithmetic)? Note that, strictly speaking, the results obtained apply only at the **nominal temperature**, as this was held constant for the 4 data points used to obtain them. Note also that the negative control results are crucial for estimating this interactive effect, which if non-negligible, may play a crucial role in impeding reproducibility.

Next consider the factorial design. There are now 4 pairs of points, the 4 vertical sides of the 2 squares, that measure the batch effect at the 4 different combinations of temperature and presence/absence of M. Note that each of these pairs can be considered a legitimate miniature OFAT: only batch changes between the two points, while the temperature and molecule status are fixed. Hence, this gives 4 differences that measure batch effect that can then be averaged over the full range of temperatures, not just the nominal. This makes results more *robust*. Obviously, twice as many differences as the OFAT provides twice the information with the same overall experimental effort.[2] Similar comments apply to the potentially important *batch*M* interaction.

How about the temperature effect? First, assuming that it is reasonably linear (possibly near flat) over the small range investigated, it can be shown that for the OFAT design only the 2 differences in results between the low and high temperatures on the bottom horizontal of each square provide information on it – the center value only helps estimate the overall average concentrations with and without M. As before, for the factorial design one can use 4 such concentration differences from all 4 horizontal sides – all little OFATs – to obtain more precise and robust information on both the average temperature effect and the *Temp*M* interaction than the OFAT provides.

Finally, because M-absent controls were run in both designs, the average effect of M over the temperatures/batches combinations tested can be computed in both designs as the average of the 4 differences between the corresponding points of the two squares. Nevertheless, the factorial design would tend to give more robust information, because the M effect is considered over a wider range of conditions where both temperature and batch were simultaneously changed. This is why Youden advocated such approaches over 50 years ago!

These calculations demonstrate another powerful feature of factorial designs: all of the design points contribute (typically equal) information to several questions that can *all* be more precisely answered without fear of confusion. More germane to this discussion, factorial designs simultaneously allow for assessment of the experimental response(s) of interest (here, the effect of organic molecules on neurons) *and* of systematic changes in experimental components that might make it difficult to reproduce those assessments. As a result, the factorial design strategy can help yield both more comprehensive *and* more robust results.

---

[2] Although the quality of this information as usual depends crucially on experimental conduct: e.g. was treatment order randomized; were measurements made on anonymized samples to avoid unconscious bias; etc.?

We now consider a more complicated, though still artificial, example that shows how these ideas can be extended to more realistic, complex situations.

# An extended, constructed example

## Preliminaries

To show how more variables can be studied, we consider a constructed example with a single "response" of interest, an experimental factor whose effect on the response is being researched, and five, not just two, additional systematic experimental factors. As Lithgow et. al. showed, there is usually no scarcity of such potential sources of systematic variation in even relatively "routine" experiments. For example, in a study of how the amount of doping of a substrate affects its measured electrical resistance, the exact temperature/time profile under which the doping is done, the lots and/or suppliers of substrates and or dopants, the experimental ovens used, and even the equipment/lab used to make the resistance measurements might be worth considering. In a preclinical study of the effect of a food additive on weight gain in mice – defined as amount of weight gained after 30 days – the age at which mice begin the experiment, the location of the cage in the lab, number of mice per cage, size of cage, type of bedding materials, and light/dark schedules are just some of the experimental factors that could affect mouse feeding behavior.

To emphasize the underlying concepts and remain agnostic to such details – although in practice, these details would be critical – we shall simply label the measured response of interest as "Resp", the research factor under study as "X", and the five experimental factors that might affect reproducibility as "A" - "E". This gives six factors in all whose possible effect on Resp is to be investigated..

Before proceeding with the example, it's worthwhile stepping back to provide a high level overview of the strategy. Again, the goal here is not to gain a detailed understanding of how either X or A-E influence Resp. That requires far too large an experiment. Rather, it is to gain a rough idea of X's effect, if any, and to determine whether experimental factors that shouldn't change that effect nor have any of their own behave as expected. This kind of study in which the goal is to gain critical "quality control" information with minimal effort is known as a "screening experiment" (e.g. https://www.itl.nist.gov/div898/handbook/pri/section3/pri3346.htm ). As mentioned previously, these have been studied and developed over approximately the last 100 years, so there is a large literature on them for a variety of audiences with various levels of expertise. There are also numerous software products available for implementation, many of which require relatively little technical (statistical) background.

This means that the particular design and analysis approach discussed here is just one of many that might fit the situation. Such flexibility allows the screening strategy to be adapted to the wide range of needs and constraints that scientific researchers actually face. However, it is also important to emphasize that no matter what specific approach is used, there is never a guarantee of infallibility. The extreme economy of the designs means that it is always possible for Nature to throw a curve that will yield misleading results. Again, good choice of design – and, of course, as much prior investigator insight as possible – minimizes such risks.

## Example design and analysis

We use a 12 run, 2-level, 6 factor (X and A-E) Plackett-Burman[3] (PB) design. This means that 12 combinations of the 6 factors are studied with each factor run at only 2 settings, call them "low" (L) and "high" (H) for convenience, where the L/H dichotomy used here is purely for notational convenience. For example, L may mean furnace 1, and H may mean furnace 2 in the doping study. The data were constructed so that X had a large "positive" effect on Resp, which here is defined to mean that Resp is higher at the H level then at the L level; A and B also had a positive effect, but only by about half as much as X; and there is also an interactive effect, which we shall comment on further shortly. The remaining 3 experimental factors (C-E) had negligible effects. To this underlying behavior was added some pseudorandom experimental "noise." See the appendix for coding details in the R statistical language.

Here is a table of the factor settings and constructed response..

**Table 3: Constructed Experimental Results**

| X | A | B | C | D | E | Resp |
|---|---|---|---|---|---|------|
| L | L | L | H | L | L | 91.5 |
| L | L | H | L | L | H | 85.8 |

---

[3] The names of the two researchers who developed this class of designs.

| | | | | | | |
|---|---|---|---|---|---|---|
| L | L | H | L | H | H | 91.7 |
| L | H | L | L | H | L | 95.8 |
| L | H | L | H | H | H | 95.5 |
| L | H | H | H | L | L | 93.6 |
| H | L | L | L | H | L | 97.7 |
| H | L | L | H | L | H | 99.3 |
| H | L | H | H | H | L | 114.8 |
| H | H | L | L | L | H | 104.7 |
| H | H | H | L | L | L | 116.5 |
| H | H | H | H | H | H | 118.8 |

Each row of this table represents one set of conditions – one experimental "run" – at which the experiment was conducted and response measured. The goal of the exercise is to recover the "true" experimental effects just described from the noisy data. The X effect is obvious: all the responses for the "L" setting of X were less than those at the "H" setting. But what about the remaining 5 possible experimental factors – (how) do they affect Resp? To determine this, we exploit a special property of 12 run Plackett-Burman designs: if 4 or fewer of the factors in the design have non-negligible effects (where "non-negligible" means distinguishable from background experimental "noise"), then essentially all the "important"[4] information on them can be recovered by restricting the analysis to just those and ignoring the remainder (Box and Tyssedal, 1996; Cheng, 1995). Clearly, X will be one important factor, so the essential challenge is to identify which of A-E are "non-negligible."

To do this, we employ a strategy described in greater detail and illustrated with several real examples in Coleman and Gunter (2014). In short, this strategy assumes that only a relative few (possibly none) of the factors under consideration will be the major actors (aka "effect sparsity") and that their effects can be seen primarily individually (so-called "main" effects) and secondarily in combination with other factors ("interactive" effects). These assumptions accord with usual scientific data analysis practice.

---

[4] Technically, all main effects and two factor interactions.

Based on these assumptions and the properties of such designs, a simple but effective approach [5] is just to plot the results separately for each of X and A-E and see which change the most between the L and H settings, like this:

# Figure 3:

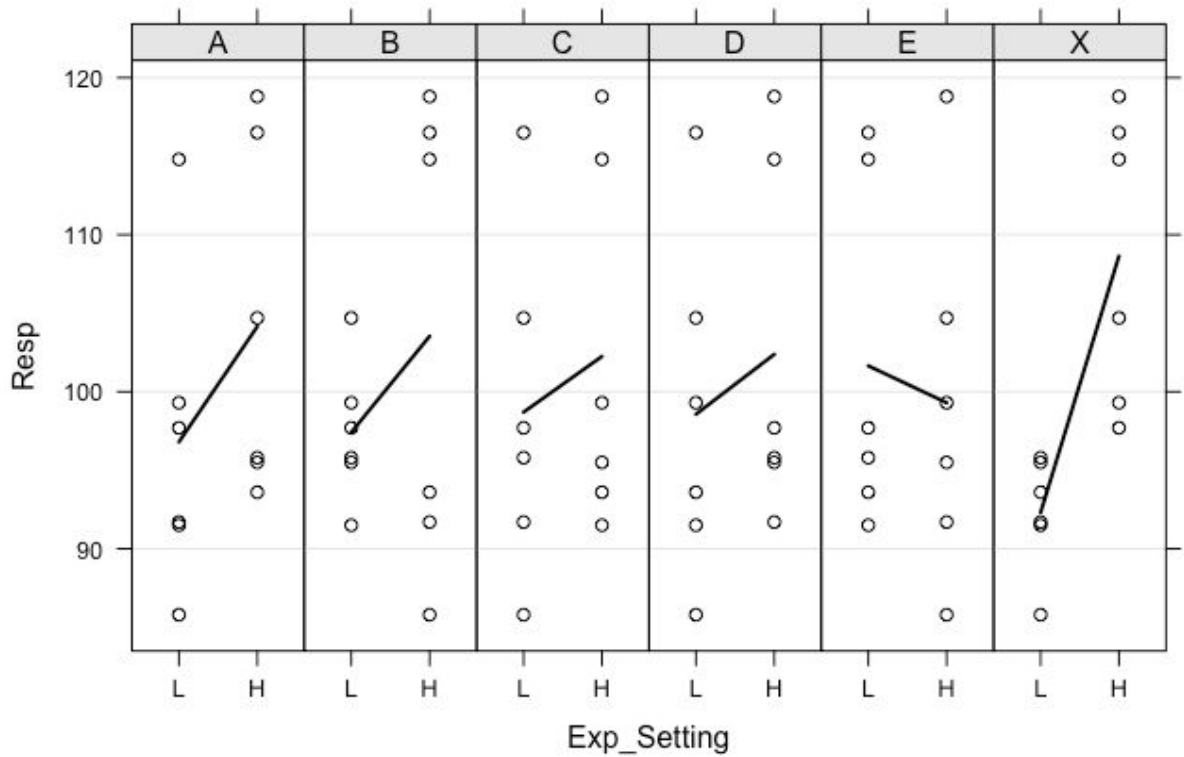

## Plots of Resp vs. Each of the Factors

The lines join the average of the six results at each setting. The X effect in the rightmost of the 6 panels acts as a benchmark for comparison. Of the five remaining experimental factors, It appears that A and B may also affect Resp, which might impact reproducibility if these factors change in other labs or attempts at replication. Note that the B panel shows something unusual: at the H level of B: the values split into two distinct groups of three, and the average sits in the middle where there are no data. This is an indication of the interactive effect of B with X, which will be interpreted more informatively below. In any case, the plots have done a reasonable job of recovering the structure that generated the data. While this may not always work so well if the data are "noisy" and the factor effects are relatively modest compared to the noise, the structure of the design[6] makes it more likely than not. In this context, it is worth mentioning that, as with the previous example in which comparisons were made between sets of four and four values, here the comparisons are between sets of six and six. So the same themes of greater robustness and more information apply.

---

[5] One could also use e.g. a main effects linear model/anova.

The next step in the analysis strategy is to examine only these most active (three) factors in greater detail. Again, we forego formal statistical methodology in favor of a graphical display, a hierarchical "structured" plot of the data. Here is the display, followed by the explanation of how to interpret it:

**Figure 4:**

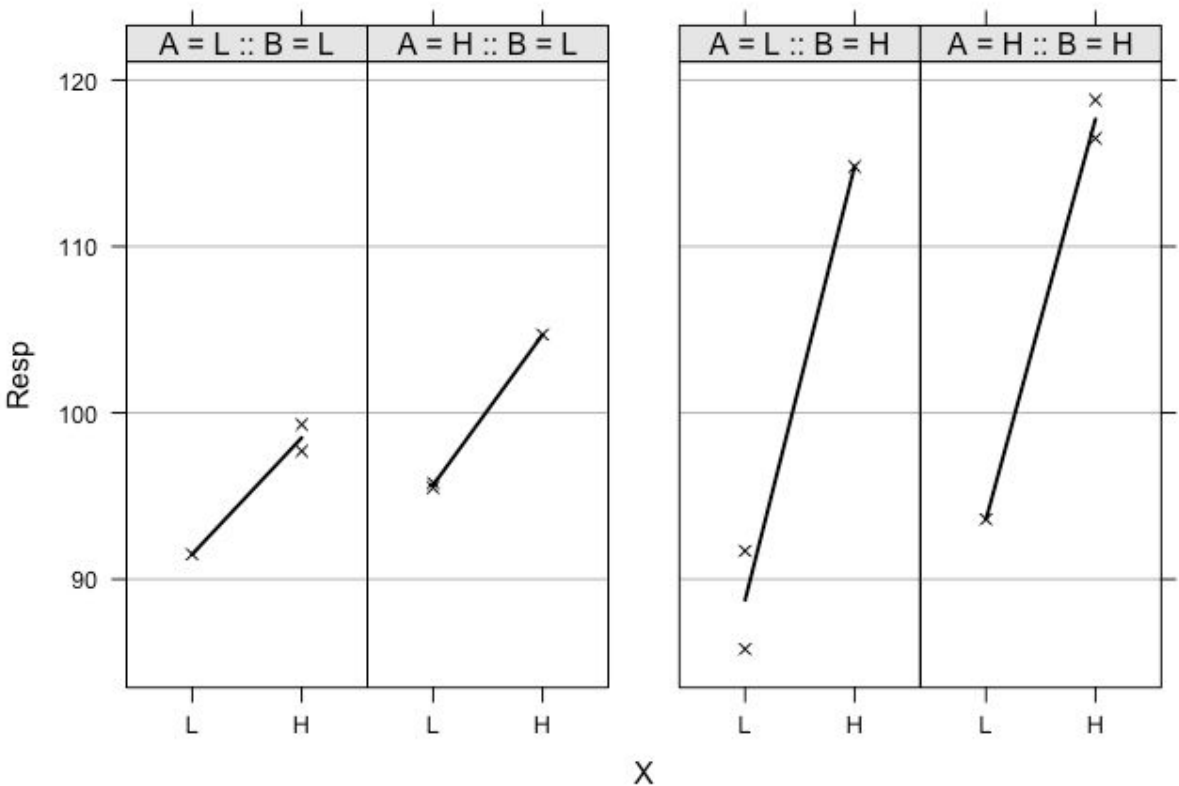

The display consists of four separate panels. Each panel shows a plot of Resp vs. X for the three data points in the panel. The four panels are laid out as two sets of two contiguous pairs. As the panel legends at the top show, A is at its "L" setting in the left panel of each pair, and at its "H" setting in the right; while the left pair of panels has B at the "L" setting and the right pair has B at "H". So, for example, in the third panel from the left in which A is at "L" and B is at "H", the three Resp values for these A and B settings in Table 1 are 85.8, 91.7, and 114.8. The first two of these occur for X = "L" and the last at X = "H". This is what is plotted in the panel.

---

[6] Specifically, its orthogonality

The effect of X is obvious as the large increase in Resp from L to H in each panel; one can also see that values at A = L are about 5 less than A=H, which might be a concern. But the B effect is what is most interesting. When B = L, changing X from L to H results in about an increase of about 8 in Resp; but when B = H, the increase in Resp is about 25, three times as large! This is an example of an interactive effect of X and B on the response: the effect of X in fact depends on what B is. Clearly, this could pose a problem for experimental reproducibility. For example, if the two B settings were actually different instruments or different lots of a reagent, this would be an indication that the experimental results could be sensitive to the particular instrument or reagent lot used. As discussed previously, OFAT experimentation cannot reveal such issues – only factorial experiments in which both the scientific change of interest, X, and the experimental factors A-E are simultaneously varied can.

## Summary

As alluded to earlier, the use of screening designs to investigate the reproducibility and robustness of experimental protocols can be readily extended to many experimental factors. For example, one could study up to five additional experimental factors (for a total of ten in all) in 12 run Plackett-Burman designs using the graphical analysis strategy demonstrated, and there are of course many other design and analysis approaches available.

To reiterate: these kinds of investigations are not intended to be comprehensive or definitive. Rather, they serve as a relatively minimalist and cheap way to investigate potential obstacles to experimental reproducibility. They should be considered as a form of experimental quality control whose purpose is to effectively exercise the experimental environment in ways that could occur in practice and raise flags for further detailed investigation if problems are found. Conversely, when such stress tests do not throw up any flags, this should provide greater assurance of the validity and replicability of scientific findings.

For implementation of these strategies, there are a host of both free (including open source) and commercial software available to handle the technical details of design and analysis and to provide any additional flexibility required to deal with any experimental constraints that may be encountered in practice. As noted previously, in practice, the limiting consideration is almost always the ability of experimenters to manage the complicated logistics required to carry out such multi-factor experiments, not the availability of designs, analytical strategies, or computational tools.

For these reasons, there appear to be no technical obstacles to the greater use of factorial designs to improve experimental robustness and reproducibility. When Youden first proposed these ideas decades ago, one might have legitimately complained about the time and effort required to learn the methodology for manual calculation or to produce informative graphical displays, which required expertise beyond the training of most scientists or engineers. The software has long since evaporated such difficulties. So it seems fair to say that the primary hindrance today is largely educational and cultural: too many scientists are either unaware of or uncomfortable with factorial designs and the greater power and flexibility that they offer. Consequently, the outmoded and inefficient OFAT paradigm still predominates despite its flaws. This needs to change.

**ACKNOWLEDGMENT**

I would like to thank my colleague Chris Tong for his encouragement and advice. His perspective and feedback were invaluable, and his patience in critiquing previous versions of this missive boundless.

# Appendix

The results for the extended example were generated in version 3.3.3 of the R statistical software( R Core Team, 2017. R: A language and environment for statistical computing. R Foundation for Statistical Computing, Vienna, Austria. https://www.R-project.org/ ) as follows: The "L" and "H" factor designations in Table 1 were first converted to numeric -1 and 1 respectively. Call the resulting table "ex"; it is a so-called "data frame" in R.  The Resp column was then generated by the R code:

```
set.seed(3001)
Resp <- round(with(ex, X*10+A*3 +2*B+ X*B*3 + rnorm(12, mean = 100, sd=3)),1)
```